\documentclass[useAMS,usenatbib]{mn2e}
\usepackage{graphicx}
\usepackage{color}

\title[A compact filament near the Galactic Centre]{Identification of a high-velocity compact nebular filament 2.2 arcsec south of the Galactic Centre}

\author[Steiner et al.] {J.~E.~Steiner,$^1$\thanks{E-mail:
steiner@astro.iag.usp.br}
R.~B.~Menezes$^1$ and Daniel~Amorim$^1$ \\
$^{1}$Instituto de Astronomia Geof\'isica e Ci\^encias Atmosf\'ericas, Universidade de S\~ao Paulo, Rua do Mat\~ao 1226, \\Cidade Universit\'aria, S\~ao Paulo, SP CEP 05508-090, Brazil}

\begin{document}

\date{Submitted 2012 December 12}

\pagerange{\pageref{firstpage}--\pageref{lastpage}} \pubyear{2013}

\maketitle

\label{firstpage}

\begin{abstract}
The central parsec of the Milky Way is a very special region of our Galaxy; it contains the supermassive black hole associated with Sgr A* as well as a significant number of early-type stars and a complex structure of streamers of neutral and ionized gas, within two parsecs from the centre, representing a unique laboratory. We report the identification of a high velocity compact nebular filament 2.2 arcsec south of Sgr A*. The structure extends over ~1 arcsec and presents a strong velocity gradient of ~200 km s$^{-1}$ arcsec$^{-1}$. The peak of maximum emission, seen in [Fe III] and He I lines, is located at $d\alpha = +0.20 \pm 0.06$ arcsec and $d\delta = -2.20 \pm 0.06$ arcsec with respect to Sgr A*. This position is near the star IRS 33N. The velocity at the emission peak is $V_r = -267$ km s$^{-1}$. The filament has a position angle of PA $= 115\degr \pm 10\degr$, similar to that of the Bar and of the Eastern Arm at that position. The peak position is located 0.7 arcsec north of the binary X-ray and radio transient CXOGX J174540.0-290031, a low-mass X-ray binary with an orbital period of 7.9 hr. The [Fe III] line emission is strong in the filament and its vicinity. These lines are probably produced by shock heating but we cannot exclude some X-ray photoionization from the LMXB. Although we cannot rule out the idea of a compact nebular jet, we interpret this filament as a possible shock between the Northern and the Eastern Arm or between the Northern Arm and the mini-spiral ``Bar''.
\end{abstract}


\section{Introduction}

The central parsec of the Milky Way (MW) is a very special region of our Galaxy. It not only hosts the central supermassive black hole, associated with Sgr A*, with well determined mass \citep{ghe08, gil09}, but also contains a significant number of early-type stars, including some in the Wolf-Rayet stage.

The central cavity has a mini-spiral structure of ionized gas streamers, encircled by a circum-nuclear disk (CND) of molecular gas. Most of the ionized gas emission appears in Hydrogen and Helium recombination lines, very likely originating from ionization produced by the many young and hot OB stars; the equivalent to some $\sim 250$ O9 stars would be required to account for the emission \citep{lac91, shi94, rob96, mor96, pau04}. An anomalous structure, called mini-cavity, was identified in the mini-spiral, with a diameter of $2\arcsec$, about $3\arcsec\!\!.5$ to the southwest of Sgr A* \citep{yus89, yus90, zha91}, associated with strong [Fe III] emission \citep{eck92, lut93}. For a recent review of the literature on the Galactic centre, see \citet{gen10}.

X-ray binary transients are overabundant in the central parsec of the MW. \citet{mun05a} have found that the overabundance is at least a factor of 20, per mass, in the central parsec when compared to the abundance at a distance between 1 and 23 parsecs. It is likely that this overabundance is due to three-body interactions between binary star systems and either black holes or neutron stars that have concentrated in the central parsec through dynamical friction. This is also supposed to be the case with the formation of similar systems in the cores of globular clusters \citep{mun05a}. 

The known X-ray transient closest to Sgr A*, at only $2\arcsec\!\!.8$ south of the centre of the MW, is CXOGX J174540.0-290031, discovered in 2004 \citep{mun05a}. It was found to be an X-ray binary with an orbital period of 7.9 hr \citep{mun05a, mun05b}. From K band observations, an upper limit for any counterpart was set as $K < 16$, suggesting that it is a low-mass X-ray binary (LMXB). This X-ray transient also revealed a radio outburst \citep{bow05}. This large radio outburst, as well as its X-ray emission, suggests that this source more likely contains a black hole than a neutron star primary \citep{fen01, por05}.

In this paper we report the discovery of a high velocity compact nebular filament extending $\sim 1\arcsec$ with a strong velocity gradient that could possibly be associated with a colliding shock between the streamers of the mini-spiral.

\section{Observations and data reduction}

This paper analyzes archive data obtained with the Spectrograph for INtegral Field Observations in the Near Infrared (SINFONI) on the VLT, available in a public access data bank\footnote{http://archive.eso.org/wdb/wdb/eso/sinfoni/form}. The observations were made in K band on April 22, 2007. A fore-optics of $0\arcsec\!\!.250$ pixel$^{-1}$ was used, providing a field of view (FOV) of $8\arcsec \times 8\arcsec$. A total of four exposures of the Galactic nucleus were retrieved from the public data bank to be used in this work.

Standard calibration images of linearity lamp, distortion fibre (used in the data reduction to compute spatial distortions and perform a spatial rectification), flat lamp, arc lamp and sky field were obtained from the public data bank. Calibration images of an A0V standard star were also retrieved from the data bank. The data reduction was made with Gasgano software and included the following steps: determination of the trim, sky subtraction, correction of bad pixels, flat-field correction, spatial rectification (including correction for spatial distortions), wavelength calibration and data cube construction. At the end of the process, four data cubes were obtained, with spaxels (spatial pixels) of $0\arcsec\!\!.125 \times 0\arcsec\!\!.125$. Afterwards, with a script developed by us and using data from the A0V standard star, we applied, in an IRAF environment, telluric absorption removal and flux calibration to all the data cubes. Details of the data treatment are given in Appendix A.

\section{Data analysis}

\begin{figure*}
  \includegraphics[scale=0.25]{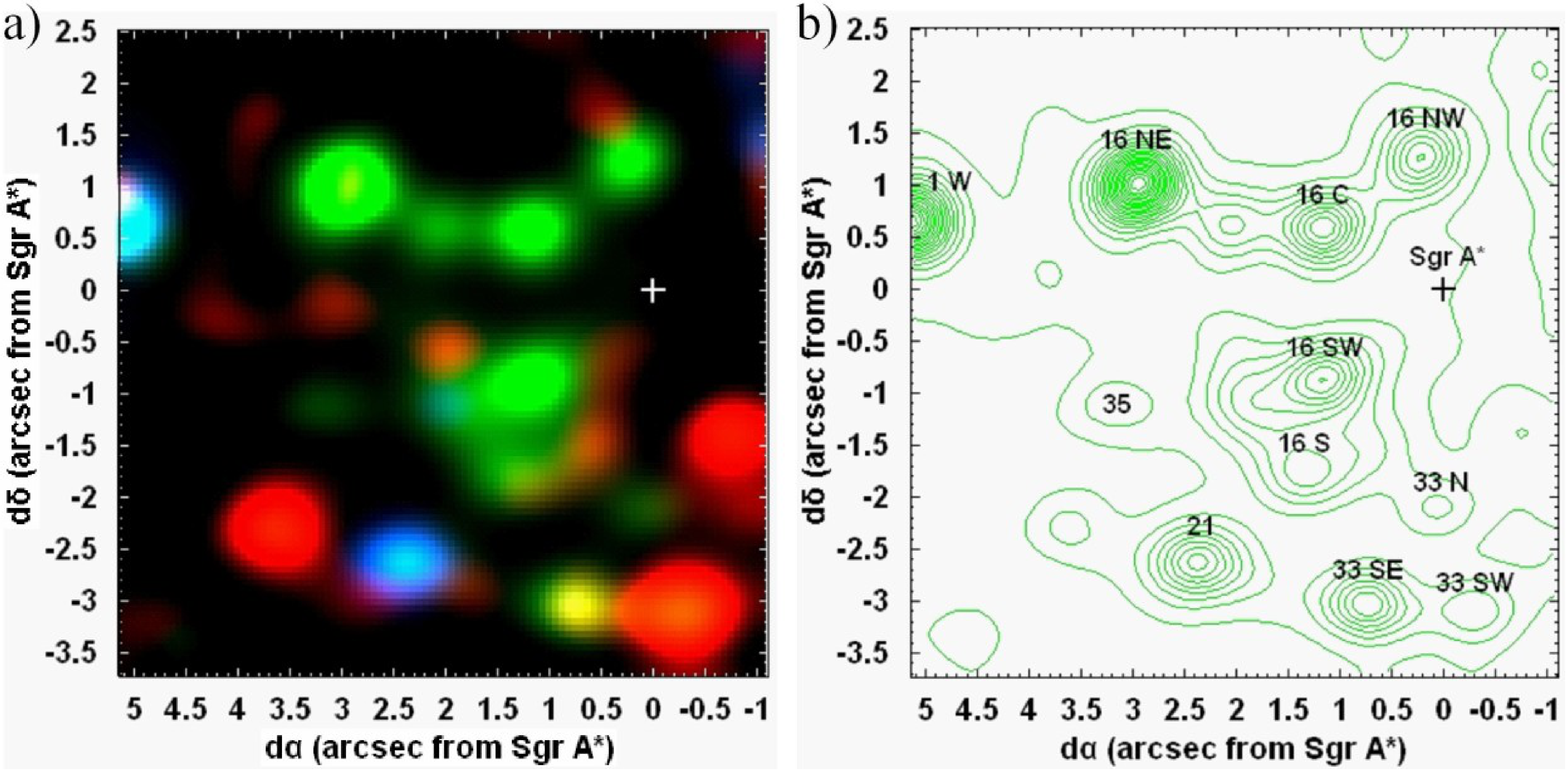}
  \includegraphics[scale=0.25]{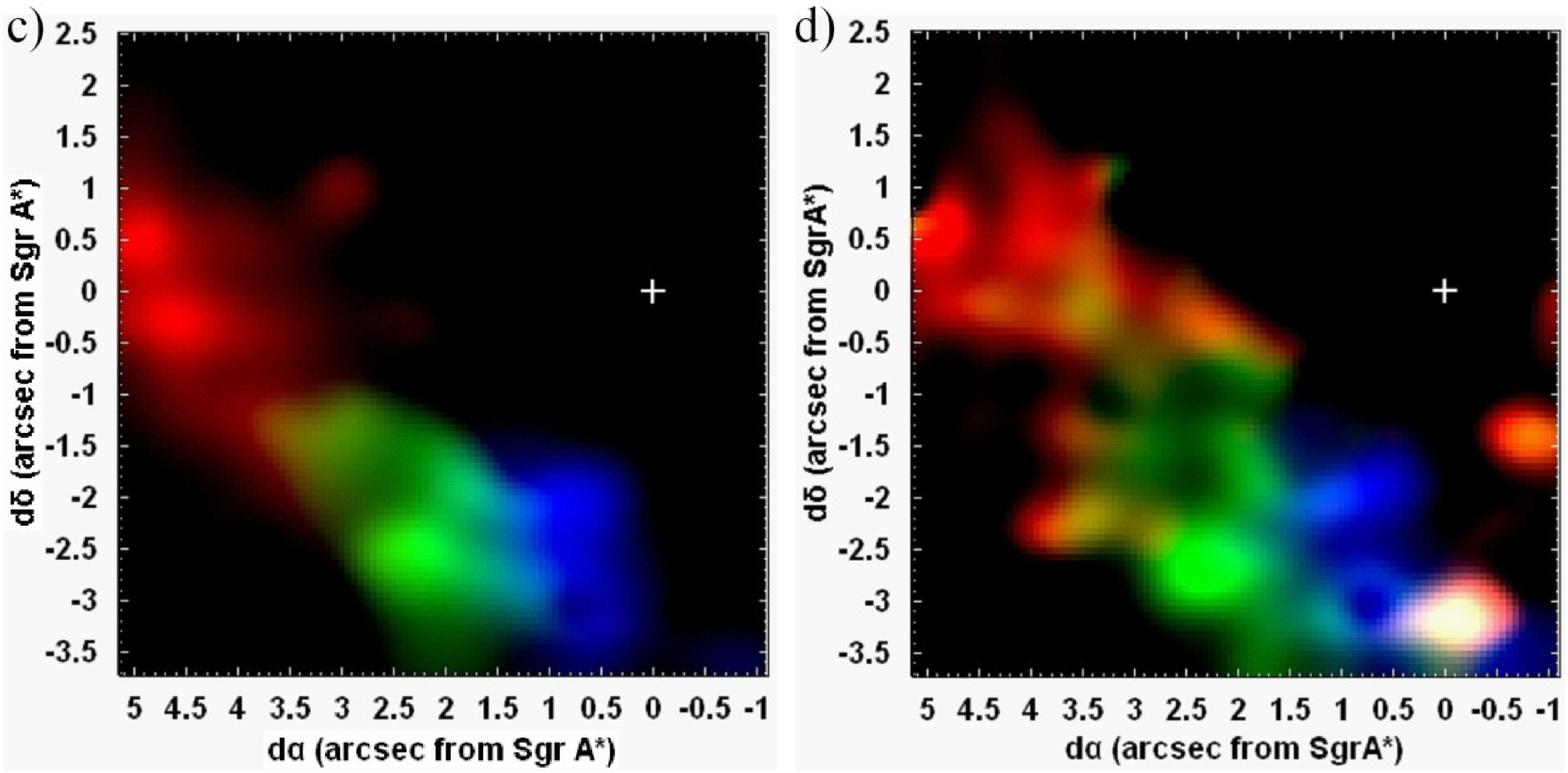}
  \includegraphics[scale=0.25]{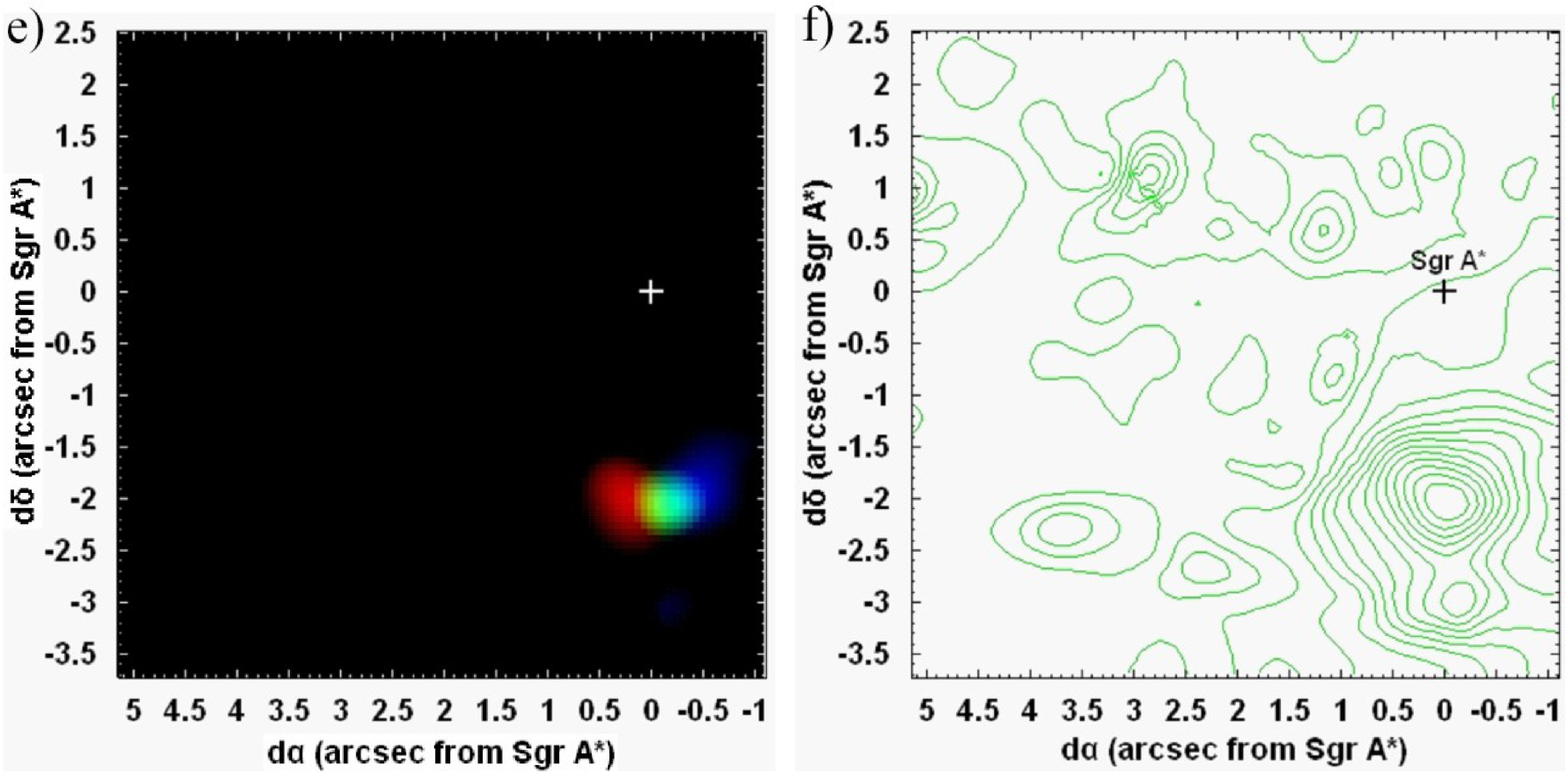}  
  \caption{a) The FOV of the SINFONI observation in pseudo colors. In green we show the young stars; in red, the old stars and in blue, the stars that have strong infrared excess. b) Contours of the stars in the FOV, with identifications. c) The FOV of the Br$\gamma$ emission line in three colors, showing the radial velocity structure of the Northern Arm. The colors blue, green and red represent the velocity ranges $-156$ km s$^{-1} \le V_r \le -73$ km s$^{-1}$, $-46$ km s$^{-1} \le V_r \le -18$ km s$^{-1}$ and $10$ km s$^{-1} \le V_r \le 93$ km s$^{-1}$, respectively. d) The same for [Fe III] $\lambda 22180 \AA$. The colors blue, green and red represent the velocity ranges $-108$ km s$^{-1} \le V_r \le 0$ km s$^{-1}$, $27$ km s$^{-1} \le V_r \le 54$ km s$^{-1}$ and $81$ km s$^{-1} \le V_r \le 189$ km s$^{-1}$, respectively. e) The filament, seen in the line [Fe III] $\lambda 22180 \AA$. The colors blue, green and red represent the velocity ranges $-379$ km s$^{-1} \le V_r \le -298$ km s$^{-1}$, $-271$ km s$^{-1} \le V_r \le -243$ km s$^{-1}$ and $-216$ km s$^{-1} \le V_r \le -135$ km s$^{-1}$, respectively. f) Contours of the image shown in e). The $+$ sign indicates the position of Sgr A*.\label{fig1}}
\end{figure*}

After the data treatment, we applied PCA Tomography (see Appendix A) again. Tomogram 1, shown in green in Figure 1-a, displays the young stellar population. Tomogram 2, shown in blue in Figure 1-a, reveals stars with strong infrared excess, presumably very young stars, like IRS 21 \citep{tan02} and IRS 1. The red component in Figure 1-a corresponds to an image of stars obtained by subtracting the edges of the CO bands from the reconstructed data cube, after suppressing eingenvectors 1 and 2 (see Steiner et al. 2009 for a description of ``feature suppression''). The stellar component revealed by Figure 1-a basically shows the known young as well as the old stellar populations. IRS 16, the closest IRS source to the Galactic centre, is resolved in the individual stars, as is IRS 33 (SW of the FOV). 

We performed spectral synthesis on the spectrum of each spaxel of the datacube analyzed here, using the Starlight software \citep{cid05}, which matches the stellar spectrum of a given object with a model corresponding to a combination of template stellar spectra from a pre-established base. In this work, we used the base of stellar spectra MILES (Medium resolution INT Library of Empirical Spectra) \citep{san06}, containing the spectra of 150 stellar populations with ages between $1.0 \times 10^6$ years and $1.8 \times 10^{10}$ years and with metallicities between 0.0001 and 0.05 (with $Z_{\sun} = 0.02$). The spectral synthesis resulted in a synthetic stellar spectrum for each spaxel of the data cube.

Before performing spectral synthesis on the data cube, however, we prepared  the spectra in two steps: correcting the interstellar extinction due to the Galaxy, using $A_V = 19$ and the reddening law of \citet{car89}; and carrying out a spectral re-sampling with $\Delta\lambda = 1\AA$ per pixel. The first step used a script written in IDL language, while the second used the task ``dispcor'', of the ``noao'' package, in IRAF environment.

\begin{figure*}
  \includegraphics[scale=0.48]{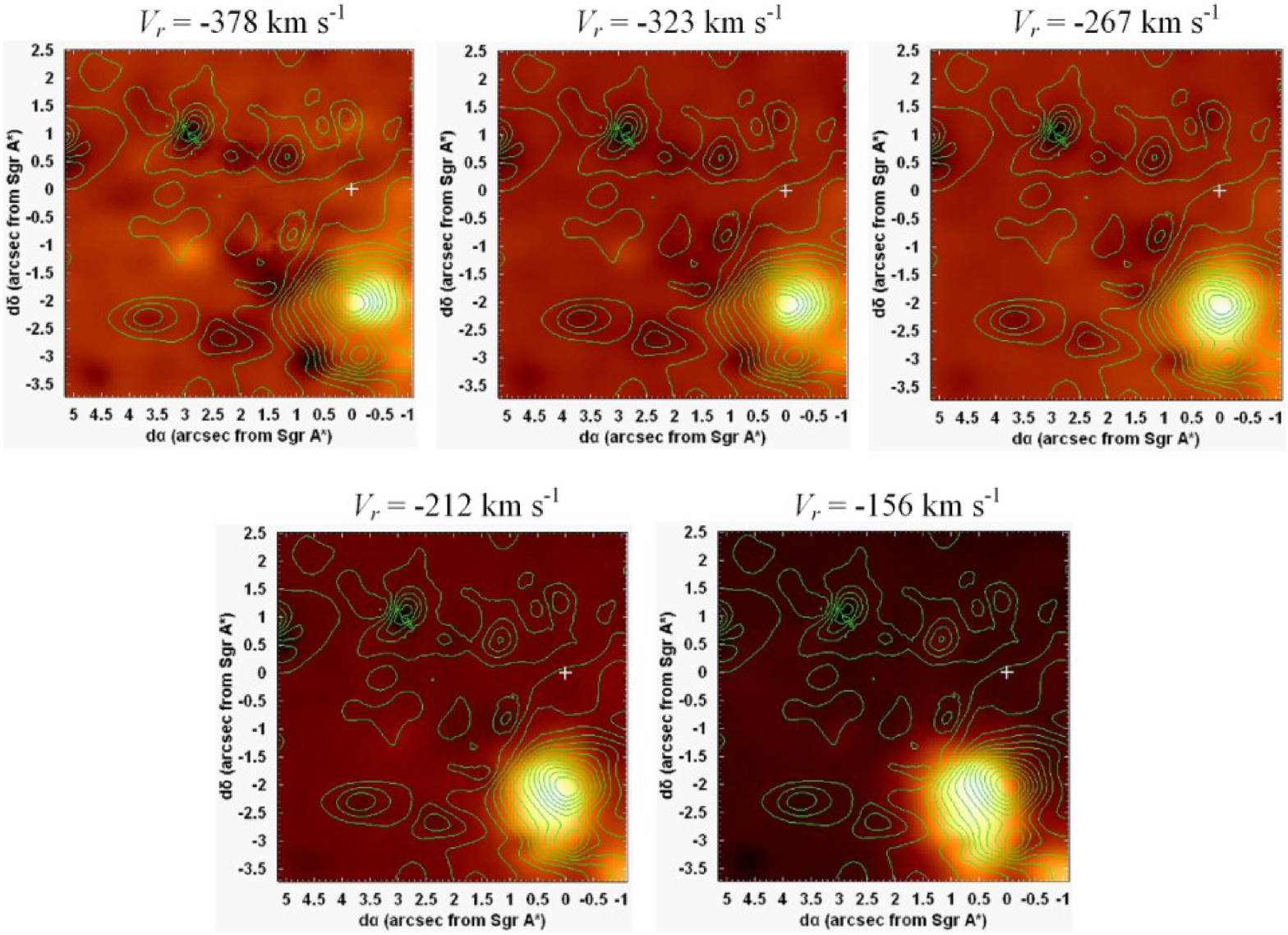}  
  \caption{Chanel maps for Br$\gamma$. The velocity of each channel is given on the top. The position of Sgr A* is marked as a $+$ sign. The contours of Figure 1-f are shown in green. We call the attention to the fact that at highest velocities, the filament is the strongest emission detected in the FOV.\label{fig2}}
\end{figure*}

In order to analyze the emission line spectrum in more detail, we subtracted the synthetic stellar spectra obtained with the spectral synthesis from the observed ones. This procedure resulted in a data cube with emission lines only. We then constructed RGB images of the emission lines Br$\gamma$ and He I $\lambda 20581\AA$ of this emission line data cube. We can see that the recombination lines Br$\gamma$ and He I $\lambda20581\AA$ are detected as stellar winds in early-type stars and, also, as ionized interstellar lines basically delineating the Northern Arm. A number of other lines are also seen in the southwest of the FOV: $\lambda21451\AA$, $\lambda22178\AA$, $\lambda22421\AA$, and $\lambda23479\AA$ are the strongest among them and are identified as [Fe III] lines (Lutz et al. 1993). We then constructed RGB images of the emission lines Br$\gamma$ and [Fe III] $\lambda22180\AA$, shown in Figure 1-c and Figure 1-d, respectively.

We identified a high velocity compact nebular filament, $2\arcsec\!\!.2$ south of Sgr A*. The structure extends over $\sim 1\arcsec$ and presents a velocity gradient of $\sim 200$ km s$^{-1}$ arcsec$^{-1}$ for He I, Br$\gamma$ and [Fe III] $\lambda 22180 \AA$ lines (Figure 1-e, and Figures~\ref{fig2} and~\ref{fig4}). The peak of maximum emission seen in the [Fe III] and He I lines is located at $d\alpha = +0\arcsec\!\!.20 \pm 0\arcsec\!\!.06$ and $d\delta = -2\arcsec\!\!.20 \pm 0\arcsec\!\!.06$ with respect to Sgr A*. This position is near the star IRS 33N, at $d\alpha = -0\arcsec\!\!.06$; $d\delta = -2\arcsec\!\!.19$ from Sgr A*. As can be seen in Figure~\ref{fig3}, the velocity at the peak, as estimated from the Br$\gamma$, He I $\lambda 20581 \AA$ and [Fe III] $\lambda 22178 \AA$ is $\sim -267$ km s$^{-1}$. The filament has a position angle of PA $= 115\degr \pm 10\degr$ (see Figures~\ref{fig2} to~\ref{fig4}).

The observed line ratios at the emission peak of the filament are (He I $\lambda 20581 \AA$)/Br$\gamma$ = $0.39 \pm 0.06$ and ([FeIII] $\lambda 22180 \AA$)/Br$\gamma$ = $0.27 \pm 0.04$.

\section{Discussion and conclusion}

The filament is positioned in one of the mini-spiral regions nearest to the Galactic centre, in the field where the Eastern and the Northern Arms of the mini-spiral cross. It is also at the eastern edge of the mini-cavity. This field shows significant blue-shifted He I emission line (Paumard et al. 2004) and also in the H92$\alpha$ and H30$\alpha$ lines \citep{zha09, zha10}.

\begin{figure*}
  \includegraphics[scale=0.28]{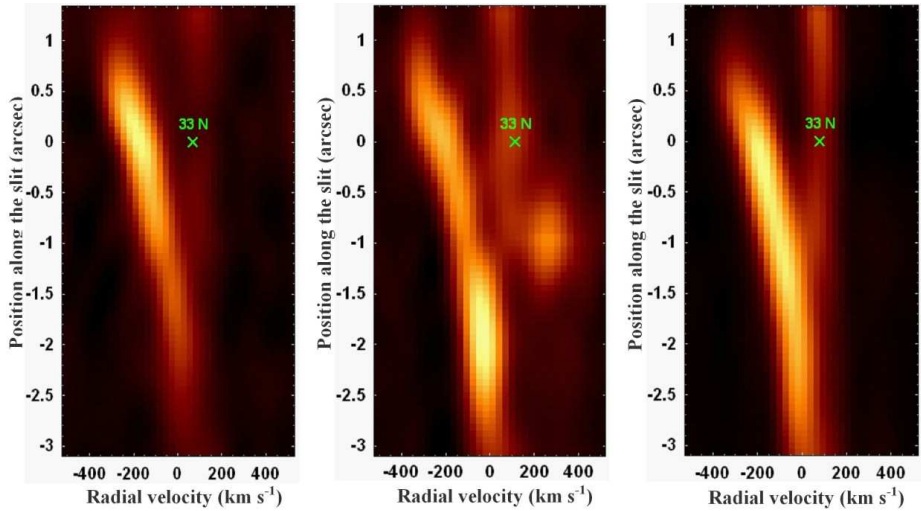}  
  \caption{Left: Pseudo-slit of the data cube along the gaseous filament, displaying spatial position versus velocity for the [Fe III] $\lambda 22180 \AA$ line. Centre: The same for the line He I $\lambda 20581 \AA$. The star at position $-1\arcsec$ and velocity $\sim +250$ km s$^{-1}$ is IRS 33SE. Note the presence of a second nebular velocity with $V_r \sim +80$ km s$^{-1}$ at the same position as the high velocity filament. This is the velocity of the Eastern Arm at that position. Right: The same for Br$\gamma$.\label{fig3}}
\end{figure*}

The filament that we see may be described as a ``jet-like'' structure, being brighter at the centre and fading towards the borders, at least as it is seen in the He I and the [Fe III] lines (see Figure~\ref{fig3}). This is in the part of the mini-spiral closest to Sgr A*. The challenge is to explain the compactness of the structure as well as its ionization (He I) and excitation ([Fe III]).

\begin{figure}
  \includegraphics[scale=0.34]{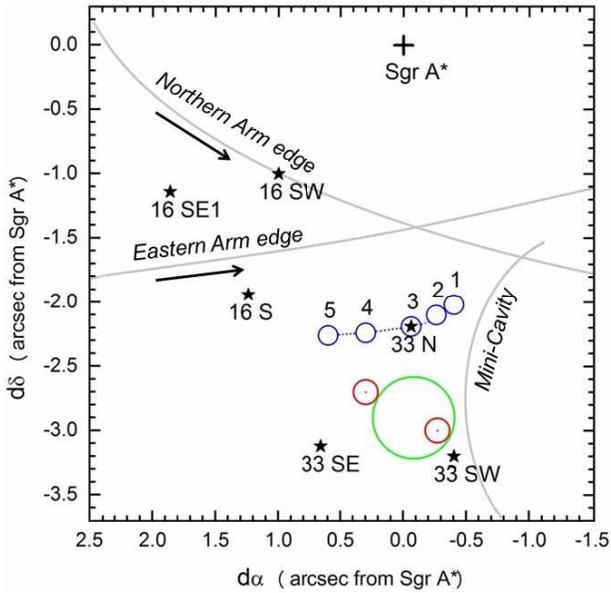}  
  \caption{The field of the nebular filaments shown at positions with distinct velocities: 1 ($-378$ km s$^{-1}$); 2 ($-323$ km s$^{-1}$); 3 ($-267$ km s$^{-1}$); 4 ($-212$ km s$^{-1}$) and 5 ($-156$ km s$^{-1}$). The northern edges of the Northern and Eastern Arm streamers, as well as of the mini-cavity, are shown. Three stars of IRS 16 as well as of IRS 33 are also shown. The position of Sgr A* is marked as a $+$ sign. The error circle of the LMXB (green) and the associated radio transients (red) are also shown.\label{fig4}}
\end{figure}

The filament overlaps spatially with the star IRS33N (see Figure 1-b and Figure~\ref{fig4}). Could this star be the origin of a jet? IRS 33N is a $K = 11.1$ star, presenting a spectral type of B0.5-1 I and has a radial velocity of $V_r = +68 \pm 20$ km s$^{-1}$ \citep{pau06} while the centre of the filament is at $V_r = -267$ km s$^{-1}$. Although close in terms of spatial projected distance, the nebular filament is kinematically decoupled from the star IRS 33N. Therefore this association can be discarded. 

\subsection{Fireworks in the mini-spiral: a shock between streamers?}

The orbits of the mini-spiral have been described as containing a one-armed spiral in a Keplerian disk (Lacy et al. 1991) that properly describes the [Ne II] $\lambda$12.8 $\mu$m. An alternative description is given by \citet{zha09}, in which the whole mini-spiral is essentially described by three streamers in elliptical orbits around Sgr A*, on the basis of H92$\alpha$ observations. The Northern and Eastern Arm streamers have high eccentricities ($e = 0.83 \pm 0.10$ and $0.82 \pm 0.15$, respectively) and cross near the periastron. The formal solution provides a projected crossing between the two orbits about $4\arcsec$ to the south of the Galactic centre. This solution has been disputed by \citet{iro12} who find that the one-armed spiral pattern may be a better description of the [Ne II] nebular emission. The observations of both H92$\alpha$ and the [Ne II] lines have good spectral resolution but poorer spatial resolution, of $\sim 1\arcsec\!\!.25$ to $2\arcsec$, than the data reported here. They could not have identified such a strong velocity gradient in a filament so compact. 

Perhaps the best explanation for the existence of this filament is that it is associated to a shock between streamers in the mini-spiral. In the one-arm pattern, the position of the filament is where the one-armed spiral ends; that could also be seen as the crossing point between the Northern Arm and the Bar. In the three ellipses model, the filament is near the position where the projected orbits of the Northern and the Eastern Arm streamers cross. It is perhaps not a coincidence that the position angle of the filament (PA $= 115\degr \pm 10\degr$) is similar to that of the Eastern Arm and also of the Bar. Perhaps the most challenging aspect of this interpretation is how to explain the large velocity gradient of $\sim 200$ km s$^{-1}$ arcsec$^{-1}$. It probably has to do with the proximity of the supermassive black hole and its steep gravitational potential gradient.

The blue arm of the filament points toward the mini-cavity. This structure was modeled as a bubble of ionized hot gas generated by a fast ($\ge 1000$ km s$^{-1}$) wind originated from one or several sources, within a few arcseconds from the mini-cavity, and blown into the partially ionized neutral gas streamers orbiting the centre of the Galaxy (Eckart et al. 1992; Wardle \& Yusef-Zadeh 1992; Lutz et al. 1993; Melia, Coker \& Yusef-Zadeh 1996). 

In fact, after computing the elliptical Keplerian orbits of the streamers, \citet{zha09} suggested that the Northern and Eastern Arms may collide in the ``Bar'' region, a few arcesconds south of Sgr A*. This is the region with the highest specific kinetic energy of both Northern and Eastern Arms (see Figure 21 in Zhao et al. 2009). According to \citet{zha09}, the high specific kinetic energy density emphasizes the possibility that the mini-cavity was created by strong shocks resulting from the collisions of the two streamers.

\subsection{The X-ray transient and origin of the nebular lines}

The peak of the emission of the filament is located $0\arcsec\!\!.7$ north of the binary X-ray transient CXOGX J174540.0-290031, a LMXB with an orbital period of 7.9 hr \citep{mun05a, mun05b}. This source is associated to one transient and one persistent albeit variable, radio sources \citep{bow05}. The X-ray transient source CXOGX J174540.0-290031 was detected on July, 5-7, 2004 and is located $2\arcsec\!\!.8$ south of Sgr A*, with an uncertainty of about $0\arcsec\!\!.3$ \citep{mun05a, mun05b}. Observations made at 43 GHz  a few months earlier, on March 28, 2004, detected two sources, named NE (northeast, the strongest one, reaching 93 mJy at 43 GHz) and SW (southwest) with a peak flux of 48 mJy. The coincidental variability of both radio and X-ray sources strongly suggests that they are associated. \citet{zha09} also report radio observations at 1.3 cm and found that the SW source is a persistent one, being detected before the transient phenomenon at the $1.1\pm0.1$ mJy level. The NE component, not detected in previous observations, was detected after the transient. These authors interpret the NE component as a hot spot associated with the head of the jet ejected from the persistent SW radio-variable source. If so, then the stellar counterpart of the X-ray binary is probably near the star IRS 33SW, close to the X-ray error box (Figure~\ref{fig4}). The position angle of the radio jet is PA $\sim 60\degr$, quite distinct from the position angle of the nebular filament. As the filament is at $\sim 0\arcsec\!\!.7$ from the X-ray source and has a distinct position angle, we may rule out any direct association between the X-ray binary and the origin of the nebular filament.

The SINFONI observations (April 22, 2007) were made a few years after the radio detection (March 28, 2004) and also after the X-ray transients detections (July 5-7, 2004). The nebular filament is characterized by a spectrum that is quite distinct from other gas emitting regions in the FOV. Br$\gamma$ and He I $\lambda 20581 \AA$ are the strongest lines. Other gas emitting regions in the FOV are probably photoionized by the young stellar population (such as in IRS 16) that produces UV photons capable of ionizing H and He (Genzel et al. 2010). Such UV photons, because of their low energy, do not heat the nebulae and some 250 O9 equivalent stars are necessary to account for the emission \citep{zha10}. The vicinity of the X-ray source is the only region in the FOV that presents such strong forbidden lines. But shock waves are also efficient in heating gas and enhancing forbidden line emissions. The field of IRS 33 is the hottest in the mini-spiral, as measured from the H30$\alpha$ and H92$\alpha$ \citep{zha09, zha10}, with a temperature of $13000 \pm 3000$ K and an electron density of $\sim 10^5$ cm$^{-3}$. The temperature of 13000 K is enough to excite the [Fe III] emission. It is not enough to ionize the gas to the Fe$^{++}$ phase. This requires a temperature of $T > 18000$ K \citep{ost06}. But the beam size of the radio observations is $\sim 2\arcsec$; thus a compact structure with $T > 18000$ K could certainly exist. For collisional ionization of helium, the temperature must be even higher:  $T > 30000$ K. We conclude that shock heating could both produce and excite the Fe$^{++}$ ion, but not ionize He. Therefore, a combination of non-shock ionization mechanism plus shock heating seems to be required.

Could the X-ray transient be, at least partially, responsible for photoionizing the filament? Hard X-rays are efficient in photoionizing gas and heating it at the same time \citep{hal83}, enhancing the emission from low ionization nebular species. However, the ionization parameter $U$ must be $U > 10^{-2}$ in order to produce high ionization species and $U \sim 10^{-3}$ in order to produce low ionization species \citep{fer83}. In the present case, the transient X-ray luminosity was $L_x \sim 4 \times 10^{34}$ erg s$^{-1}$.  However, at the time of the transient, diffuse X-ray emission was observed $ \sim 2\arcsec - 3\arcsec$ southwest of the source. This suggests that the real X-ray luminosity was $L_x \ge 2 \times 10^{36}$ erg s$^{-1}$ \citep{mun05b}. The discrepant X-ray luminosity seen directly is probably due to the fact that the X-ray binary is seen nearly edge-on.

Assuming that the x-ray luminosity is $L_x \sim 2 \times 10^{36}$ erg s$^{-1}$ we obtain an ionization parameter of $U \sim 5 \times 10^{-6}$ at the filament; this is too low to produce a significant ionization effect. We conclude that the X-ray binary has no ionization or heating effect on the filament unless the source radiates soft X-rays near the Eddington limit. This could not have been directly observed due to the strong interstellar absorption and would produce $U \sim 10^{-2}$ at the filament and could be responsible for the strength of the [Fe III] emission and, perhaps, of He I.

Further studies are required to elucidate the nature of this filament and the role of the LMXB in its neighborhood; it is important to observe the FOV with better spatial resolution.

\subsection{Summary}

We have analyzed a data cube centered $\sim 2\arcsec$ southeast of Srg A*, obtained with the SINFONI/VLT IFU spectrograph and studied the kinematics of the gaseous ionized emission in the FOV. Our main findings are:

\begin{itemize}

\item We identified a compact nebular filament, $2\arcsec\!\!.2$ south of Sgr A*. The structure extends over $\sim 1\arcsec$ and presents a velocity gradient of $\sim 200$ km s$^{-1}$ arcsec$^{-1}$. There is a peak of maximum emission seen in [Fe III] and He I lines, located at $x = +0\arcsec\!\!.20 \pm 0\arcsec\!\!.06$ and $y = -2\arcsec\!\!.20 \pm 0\arcsec\!\!.06$ with respect to Sgr A*. This position is located at $\sim 0\arcsec\!\!.3$ east of the star IRS 33N. This emission has the most negative velocity in the FOV we have analyzed.

\item The velocity at the emission peak, as estimated from the Br$\gamma$, He I $\lambda 20581 \AA$ and [Fe III] $\lambda 22178 \AA$ is $\sim -267$ km s$^{-1}$. The nebular filament has a position angle of PA $= 115\degr \pm 10\degr$, similar to that of the Eastern Arm at that position and to the position angle of the mini-spiral Bar. Although close, the nebular filament is, kinematically decoupled from the star IRS 33N, which has a velocity of $V_r = +68 \pm 20$ km s$^{-1}$.

\item The peak position is also located $0\arcsec\!\!.7$ north of the binary X-ray transient CXOGX J174540.0-290031, a LMXB with an orbital period of 7.9 hr. This source is associated with one transient plus one persistent and variable radio sources.

\item The [Fe III] line emission is strong in the filament and its vicinity. These lines are probably excited by shock heating.  Ionization/heating by soft X-ray photons from the nearby LMXB is possible, although unlikely, because of the high luminosity required.

\item Although we cannot rule out the possibility that we are dealing with a jet-like structure, we propose that the filament may represent a shock from the collision between the Northern and the Eastern Arm streamers around the centre, or perhaps, between the Northern Arm and the Bar. 

\end{itemize}

\section*{Acknowledgments}

This research has been supported by the agencies CNPq and FAPESP. The data presented herein were obtained with the SINFONI spectrograph on the VLT/ESO available on the ESO public data bank.

\appendix

\section[]{Data treatment}

After the data reduction, a correction of the differential atmospheric refraction was applied to all data cubes, using an algorithm developed by our group. It is important to mention, however, that the relative shifts observed in the images of all data cubes are not compatible with what is expected for the differential atmospheric refraction effect. Therefore, we believe that these relative shifts are, at least partially, a consequence of some instrumental effect and not entirely due to the differential atmospheric refraction. In order to combine into one the four data cubes obtained after the correction of the atmospheric differential refraction, we divided the data cubes in two groups with three data cubes each. Naming the data cubes 1, 2, 3 and 4 (according to the order of the observations), the division in two groups was the following: \\
\\
Group 1: 1, 2 and 3 \\
Group 2: 2, 3 and 4\\
\\
We then calculated a median of the data cubes of each group, which resulted in two data cubes at the end of the process. Finally, we calculated the average of the two data cubes and obtained the final, combined data cube.

By performing a spatial re-sampling to all the images of the data cube, we obtained spaxels of $0\arcsec\!\!.0625 \times 0\arcsec\!\!.0625$. This procedure does not change the spatial resolution of the observation, but sharpens the contours of the structures in the images. The disadvantage of such spatial re-sampling is that it introduces high spatial frequency components in the images (usually in the form of small vertical and horizontal stripes). The high spatial frequency components, however, are not a serious problem because, after the spatial re-sampling, a Butterworth spatial filtering was applied to all the images of the data cube, to remove high spatial frequency noises, including the high frequency components introduced by the spatial re-sampling. The steps of this process were as follows: calculation of the Fourier transforms ($F(u,v)$) of all the images of the data cube; multiplication of the Fourier transforms $F(u,v)$ by the image corresponding to the Butterworth filter ($H(u,v)$); calculation of the inverse Fourier transforms of all the products $F(u,v) \times H(u,v)$; extraction of the real part of the calculated inverse Fourier transforms. The Butterworth filter used in this case corresponds to the product of two identical circular filters, with orders $n = 2$ and cutoff frequencies along the horizontal and vertical axis equal to 0.27 Ny (Nyquist frequency).

We detected the presence of a probable instrumental ``fingerprint'' in the data cube, after the Butterworth spatial filtering, which appeared in the data cube as a large horizontal strip in the images and also had a low frequency spectral signature. In order to remove it, we applied the PCA Tomography technique \citep{ste09} to the data cube. PCA transforms data expressed originally in correlated coordinates into a new system of uncorrelated coordinates (eigenvectors) ordered by principal components of decreasing variance. PCA Tomography consists in applying PCA to data cubes. In this case, the variables correspond to the spectral pixels of the data cube and the observables correspond to the spaxels of the data cube. Since the eigenvectors are obtained as a function of the wavelength, their shape is similar to spectra and, therefore, we call them eigenspectra. On the other hand, since the observables are spaxels, their projections on the eigenvectors are images, which we call tomograms.  The simultaneous analysis of eigenspectra and tomograms allows one to obtain information that, otherwise, would possibly be harder to detect. Using PCA Tomography, we were able to identify and remove the instrumental fingerprint of the data cube analyzed here (Steiner et al., in preparation).

Finally, a Richardson-Lucy deconvolution \citep{ric72, luc74} was applied to all the images of the data cube, using a synthetic Gaussian PSF. The use of a Gaussian PSF was possible in this case because we verified that the original PSF of this observation had an approximate Gaussian shape. The final data cube obtained after the full data treatment has a point-spread function (PSF) with FWHM $\sim 0\arcsec\!\!.56$ and a spectral resolution of $R \sim 5875$.

\end{document}